# Suppression of Phase Separation in LiFePO$_4$ Nanoparticles During Battery Discharge

*Peng Bai,*[†,§] *Daniel A. Cogswell*[†] *and Martin Z. Bazant*\*[,†,‡]

[†]Department of Chemical Engineering and [‡]Department of Mathematics, Massachusetts Institute of Technology, 77 Massachusetts Avenue, Cambridge, Massachusetts 02139, USA

[§]State Key Laboratory of Automotive Safety and Energy, Department of Automotive Engineering, Tsinghua University, Beijing 100084, P. R. China

\*Corresponding author: Dr. Martin Z. Bazant. Tel: (617)258-7039; Fax: (617)258-5766; Email: bazant@mit.edu

August 8, 2011

ABSTRACT: Using a novel electrochemical phase-field model, we question the common belief that Li$_x$FePO$_4$ nanoparticles separate into Li-rich and Li-poor phases during battery discharge. For small currents, spinodal decomposition or nucleation leads to moving phase boundaries. Above a critical current density (in the Tafel regime), the spinodal disappears, and particles fill homogeneously, which may explain the superior rate capability and long cycle life of nano-LiFePO$_4$ cathodes.

KEYWORDS: Li-ion battery, LiFePO$_4$, phase-field model, Butler-Volmer equation, spinodal decomposition, intercalation waves.

*Introduction.* – LiFePO$_4$-based electrochemical energy storage[1] is one of the most promising developments for electric vehicle power systems and other high-rate applications such as power tools and renewable energy storage.[2,3] Nanoparticles of LiFePO$_4$ have recently been used to demonstrate ultrafast battery discharge[4] and high power density carbon pseudocapacitors.[5] However, the design of these systems remains both challenging and controversial due to the poor understanding of lithium intercalation dynamics in nanoparticles with highly anisotropic transport,[6,7] size-dependent diffusivity,[8] and a strong tendency to separate into Li-rich and Li-poor phases.[9-11] The fact that reported values of the diffusivity and exchange current for LiFePO$_4$ vary by orders of magnitude [6,8,12-16] indicates fundamental uncertainty in the kinetic processes governing battery operation and highlights the need for models that take the unique properties of LiFePO$_4$ into account.

The prevailing belief is that phase separation always occurs in Li$_X$FePO$_4$ during battery operation. This assumption is built into widely accepted porous-electrode models,[12,13,17] where each particle is modeled an isotropic sphere with a "shrinking core" of one phase displaced by a "shell" of the other phase, as suggested in the first experimental paper.[1] In contrast, phase-field models provide a more general, thermodynamically consistent treatment of nucleation and growth without resorting to the artificial placement of phase boundaries.[18,19]

Considering the excitement surrounding LiFePO$_4$ as a phase-separating cathode material, phase-field methods have received relatively little attention. Tang *et al.* modeled phase separation and crystalline-to-amorphous transformations in spherical isotropic LiFePO$_4$ particles with a phase-field model,[20] prompting experiments to seek the predicted amorphous surface layers.[21] Kao *et al.* compared the model to x-ray diffraction data and first proposed the idea of overpotential-dependent phase transformation pathways.[22] It is now becoming appreciated that models should account for the strongly anisotropic transport in crystalline LiFePO$_4$,[6,7] as well as surface reaction kinetics, which are likely to be rate limiting in nanoparticles.[23-25] Singh, Ceder and Bazant (SCB)[26] first modeled reaction-limited intercalation in LiFePO$_4$ nanoparticles by coupling an anisotropic phase-field model with Faradaic reactions on the active facet. They predicted moving phase boundaries ("intercalation waves") sweeping across FePO$_4$ planes, consistent with experimental inferences of a "domino cascade" [27] as well as the suppression of equilibrium phase separation with decreasing particle size,[28] and Tang *et al.* recently modeled the effect of coherency strain on intercalation wave structure.[29] No model, however, has yet addressed phase-separation dynamics under the experimentally relevant condition of constant applied current.

In this article, we develop a theory of reaction-limited intercalation in anisotropic nanoparticles and predict that phase separation is suppressed above a critical current. The possibility of a solid-solution pathway for LiFePO$_4$ intercalation has also been suggested by Malik *et al.*[30] based on equilibrium bulk free energy calculations, but here we show that phase separation *out of equilibrium* (i.e. during battery discharge) is controlled by *surface reactions*, which lead to unstable "quasi-solid solutions" and ultimately homogeneous filling, as current increases.

*General theory.* – We assume a regular solution model for the homogeneous free energy density,[31] and adopt a diffusional chemical potential derived from the Cahn-Hilliard free energy functional[32,33] for intercalated lithium:

$$\mu = \Omega(1-2\tilde{c}) + 2k_B T \ln\left(\frac{\tilde{c}}{1-\tilde{c}}\right) - V_s K \nabla^2 \tilde{c} = k_B T \ln a \qquad (1)$$

$\tilde{c}$ is the local filling (mole) fraction of lithium in FePO$_4$; the regular solution parameter $\Omega$ is the enthalpy of mixing per site (see supporting information); $k_B$ is Boltzmann's constant and $T$ is the absolute temperature; $V_s$ is the volume per intercalation site; $K$ is the Cahn-Hilliard gradient energy coefficient,[32] and $a = \gamma \tilde{c}$ is the local activity of lithium, scaled to unity at the voltage plateau. The factor $2k_B T$ accounts for the configurational entropy of both lithium ions and electrons.[34]

The diffusional chemical potential is appropriate for systems with a fixed number of available lattice sites. It is the free energy change per site for adding one lithium ion while consuming one vacancy, and defines the electrochemical potential of the reduced state (left hand side) of the Faradic reaction:

$$\text{LiFePO}_4 - \text{FePO}_4 \rightleftharpoons \text{Li}^+ + e^- \qquad (2)$$

The oxidized state (right hand side) consists of a lithium ion in the electrolyte and an electron that either diffuses through the solid or conducts through a metallic additive. The electrochemical potential of the oxidized state, $\mu_O = (k_B T \ln a_+ + e\phi) - e\phi_e$, consists of the chemical and electrostatic energies of Li$^+$ in the electrolyte plus the electrostatic energy of electrons. $a_+ = \gamma_+ \tilde{c}_+$ is the local activity of the Li$^+$ in the electrolyte, and $\phi$ and $\phi_e$ are the mean electrostatic potentials of ions and electrons, respectively. The local voltage drop across the interface is $\Delta\phi = \phi_e - \phi$.

Next we use transition state theory for concentrated solutions to relate the voltage to the local current density $J$ through the solid surface for the net cathodic reaction (see supporting information). We assume that the mean electrostatic energy of the activated state is equal to $\alpha$ times that of the reduced state (zero) plus $(1 - \alpha)$ times that of the oxidized state, which leads to the Butler-Volmer (BV) equation: [35]

$$J = J_0 \left[ \exp\left(-\alpha \frac{e\eta}{k_B T}\right) - \exp\left((1-\alpha)\frac{e\eta}{k_B T}\right) \right] \qquad (3)$$

where $\alpha$ is the electron-transfer symmetry factor, and $\eta = \Delta\phi - \Delta\phi_{eq}$ is the surface overpotential (the activation polarization only). The Nernst equilibrium voltage $\Delta\phi_{eq}$ and the exchange current density $J_0$ are given by:

$$\Delta\phi_{eq} = \frac{k_B T}{e} \ln\frac{a_+}{a} \qquad \text{and} \qquad J_0 = \frac{e\, a_+^{1-\alpha} a^\alpha}{A_s \tau_0 \gamma_A} \qquad (4)$$

where $A_s$ is the area of a reacting site, and $\tau_0$ is the mean time for a single reaction step. $\gamma_A$ is the chemical activity coefficient of the activated state, which we take to be $(1-\tilde{c})^{-1}$ to account for excluded volume at the reacting site.

Equations (3)-(4) represent the most general form of the BV equation for concentrated solutions. Both $\Delta\phi_{eq}$ and $J_0$ depend on the lithium concentration *and its gradients* via Eq. (1). This unusual rate dependence focuses reactions where the phase boundary meets the active facet. Our modified BV model reduces to the standard BV equation in the limit of a dilute solution ($\gamma = \gamma_+ = \gamma_A = 1$), thus correcting the SCB rate expression for consistency with statistical thermodynamics. It is straightforward to add elastic strain energy to $\mu$ or activation strain energy[36] to $\gamma_A$, and our modified BV equation (3)-(4) is a natural boundary condition for the Cahn-Hilliard equation in the solid[26] and modified Poisson-Nernst-Planck equations[37,38] or porous electrode theory in the electrolyte.[12,17,39]

*Reduced model.* – In order to capture just the essential physics of nano-LiFePO$_4$, we take a simple limit of the general theory, valid for a single nanoparticle. The active material in high-rate LiFePO$_4$ cathodes is an ultrafine powder with 30-100 nm particle sizes.[4,5] At such small length scales, the electrolyte concentration $\tilde{c}_+$ is nearly uniform and quasi-steady, since the diffusion time through the electrolyte (< 10 μs) is much smaller than through the solid (< 10 ms). For simplicity we set $a_+ = 1$ around the particle so that $\Delta\phi_{eq} = -\mu/e$ and $\eta = \Delta\phi + \mu/e$. Note that $\eta < 0$ for a cathodic reaction (insertion). Neglecting macroscopic voltage losses (electrolyte concentration polarization, separator and anode resistances, etc.), the operating voltage of our model nanoparticle battery is $V = V^\ominus + \Delta\phi$, where $V^\ominus$ is the standard potential defined by open circuit voltage plateau ($V^\ominus = 3.42$ V vs. Li metal).

Next we take into account the strongly anisotropic transport properties of the crystal, which are neglected in spherically symmetric continuum models.[12,17,20] Experiments and *ab initio* calculations have shown that Li transport in a perfect Li$_X$FePO$_4$ crystal is confined to 1D channels in the *y*-direction ($b_{pnma}$) with little possibility of transverse diffusion[6,7] (See Fig. 1a). In larger microparticles, randomly distributed defects provide sites for channel blocking and inter-channel hopping, causing diffusivity to decrease with increasing particle size.[8] Here we focus on defect-free nanoparticles and neglect the possibility of amorphization included in other isotropic particle models.[20]

Following SCB, we neglect phase separation in the y-direction and assume that the bulk concentration in each y-channel quickly equilibrates to the surface concentration, due to fast diffusion. With decreasing particle radius R, the diffusion time $\tau_d = R^2/D$ decreases as R$^2$, while the reaction time decreases only as $\tau_r = \tau_0 R/l$ (where *l* is the length of a Li site), making intercalation reaction-limited in sufficiently small particles. For an ideal crystal with R=25nm (a 50nm particle with two active surfaces) and D =10$^{-8}$ cm$^2$/sec from *ab initio* calculations,[6] the diffusion time is only $\tau_d$=0.6ms, which is much less than the shortest reported discharge time of 10 seconds.[4] Defects can lead to reduced diffusivity,[8] but only for particles larger than 100nm. Experiments cannot observe diffusion in single nanoparticles, but even with D =$10^{-13}$-$10^{-12}$ cm$^2$/s from a recent study,[13] the diffusion time for a 50 nm particle is only 1 minute.

We thus arrive at a reaction-limited model where the 3D concentration profile is represented by a 2D pattern $\tilde{c}(\tilde{x},\tilde{z},\tilde{t})$ over the active (010) facet, which satisfies the following dimensionless nonlinear PDE:

$$\frac{\partial \tilde{c}}{\partial \tilde{t}} = \tilde{J}_0 \left[ \exp(-\alpha \tilde{\eta}) - \exp((1-\alpha)\tilde{\eta}) \right] \qquad (5)$$

We scale length to the facet width $L$, voltage to the thermal voltage $k_B T/e$, and time to the mean time $N_H \tau_0$ to fill a channel of $N_H$ sites. The dimensionless exchange current density $\tilde{J}_0 = A_s \tau_0 J_0 / e$ and overpotential $\tilde{\eta} = e\eta/k_B T$ are given by:

$$\tilde{J}_0 = \tilde{c} \exp\left[ \frac{1}{2}\left( \tilde{\Omega}(1-2\tilde{c}) - \tilde{K}\tilde{\nabla}^2 \tilde{c} \right) \right] \qquad (6)$$

$$\tilde{\eta} = \tilde{\mu} + \Delta\tilde{\phi} = \tilde{\Omega}(1-2\tilde{c}) + 2\ln\left( \frac{\tilde{c}}{1-\tilde{c}} \right) - \tilde{K}\tilde{\nabla}^2\tilde{c} + \Delta\tilde{\phi} \qquad (7)$$

where $\tilde{K} = V_s K / k_B T L^2$ is the dimensionless gradient energy coefficient that penalizes sharp interfaces, introducing interfacial energy and setting the diffuse interface width.

Since LiFePO$_4$/FePO$_4$ phase boundaries are most likely to be aligned with the *yz*-plane (*bc-plane*) for minimum strain,[40] we further simplify Eqs. (5)-(7) to a one-dimensional PDE for the depth-averaged filling fraction $\tilde{c}(\tilde{x},\tilde{t})$ for a particle of length $\tilde{L}=1$ in the direction of phase separation. The total current is an integral over the active facet area $A$, $I = \int_A J \, dA$, and takes the dimensionless form:

$$\tilde{I} = \frac{\tau_0 I}{e N_A} = \int_0^1 \frac{\partial \tilde{c}}{\partial \tilde{t}} d\tilde{x} \qquad (8)$$

The exchange current is $I_0 = N_A e/\tau_0$, and $N_A = A/A_s$ is the number of active surface sites (or lithium channels). Under galvanostatic conditions, equation (8) is an integral constraint that implicitly determines the voltage $\Delta\tilde{\phi}(\tilde{t})$.

Equations (5)-(8) describe a new class of electrochemical phase transformation kinetics. While Cahn-Hilliard[32] and Allen-Cahn[41] kinetics describe phase separation with and without local conservation, respectively, our model describes phase separation under a boundary integral constraint. It applies to systems in contact with a reservoir of mass at fixed chemical potential with an imposed total flux and assumes a nonlinear flux-potential (current-voltage) relation, suitable for electrochemical reactions. The model provides a simple paradigm to understand non-equilibrium pattern formation driven by an applied voltage or current.

*Phase separation at constant voltage.* – The condition of thermodynamic equilibrium in Eq. (7) is $\tilde{\eta} = \Delta\tilde{\phi} + \tilde{\mu} = 0$, as illustrated in Fig. 2. Applying a $\Delta\tilde{\phi}$ has the effect of raising the energy of one phase relative to the other phase, and the intersections of constant $\Delta\tilde{\phi}$ with the homogeneous chemical potential curve $-\tilde{\mu}_h$ correspond to equilibrium states. For large applied voltages (i.e. $\Delta\tilde{\phi} > \Delta\tilde{\phi}^{max} = -\tilde{\mu}_h^{min}/e \approx 1.54$ or $\Delta\tilde{\phi} < \Delta\tilde{\phi}^{min} = -\tilde{\mu}_h^{max}/e \approx -1.54$), there is only one solution $A_2$ (or $B_2$), corresponding to a stable Li-poor (or Li-rich) composition. At the critical voltage $\Delta\tilde{\phi}^{max}$ or $\Delta\tilde{\phi}^{min}$ a second solution appears, corresponding to a spinodal composition that is linearly unstable to composition fluctuations. In a closed system at zero current, the instability leads to stable phase separation with constant mean composition. In an open system at constant voltage, transient phase separation can occur, but the only stable state is a homogeneous solution. As shown below, the most unstable wavelength for spinodal decomposition is set by the particle size, similar to Allen-Cahn dynamics.[41]

Three solutions appear for small applied voltages where $\Delta\tilde{\phi}^{min} < \Delta\tilde{\phi} < \Delta\tilde{\phi}^{max}$. The two extreme solutions $A_1$ and $B_1$ are linearly stable. The middle solution is unstable, lying within the spinodal region. Due to the applied voltage, one outer solution ($A_1$) has a higher energy than the other ($B_1$), which is metastable. Phase transformation can be triggered with an interface that transforms the metastable phase into the stable phase. Focusing the reaction at the phase boundary causes it to propagate across crystal planes as an "intercalation wave",[26] as illustrated in Fig. 1, with a velocity, $\tilde{v} \sim -\tilde{\lambda}\Delta\tilde{\phi}/2$, scaling as the applied voltage (see supporting information). Stable phase separation ($v=0$) occurs at the plateau voltage where $\Delta\tilde{\phi} = 0$.

*Phase separation at constant current.* – To investigate the feasibility of phase separation during battery operation, we consider a homogeneous system at concentration $\tilde{c}$ under an applied current $\tilde{I}$. A miscibility gap cannot be defined since the system is out of equilibrium, but there is a linearly unstable spinodal region. This driven instability at constant total flux is the non-equilibrium thermodynamic origin for phase separation in a homogeneously filling state. From Eqs. (5)-(8), the current-voltage relation is:

$$\Delta\tilde{\phi} = \Delta\tilde{\phi}_{eq}(\tilde{c}) + \tilde{\eta} = -\tilde{\mu}(\tilde{c}) - 2\sinh^{-1}\left(\frac{\tilde{I}}{2\tilde{J}_0(\tilde{c})}\right) \qquad (9)$$

The first term is the (homogeneous) open circuit voltage, and the second is the overpotential, which is derived from BV equation (3) with $\alpha = 1/2$. The battery voltage is the change in total free energy $G$ per electron transferred from the anode to the cathode, which has the dimensionless form $\Delta\tilde{\phi} = -d\tilde{G}/d\tilde{c}$. By applying a secant construction[18] to $\tilde{G}(\tilde{c})$ and neglecting finite-size effects, we find that the system undergoes spinodal decomposition if $d^2\tilde{G}/d\tilde{c}^2 < 0$, or $d\Delta\tilde{\phi}/d\tilde{c} > 0$. The spinodal region has a novel dependence on the applied current via Eq. (9), and is captured by the stability boundary in Fig. 3.

During battery discharge, $\tilde{I} > 0$ and the spinodal range shrinks as the concentration-dependent overpotential overcomes the solid-solution voltage barrier, $\Delta\tilde{\phi}_{ss} = \Delta\tilde{\phi}^{max} - \Delta\tilde{\phi}^{min}$. Above a critical current $\tilde{I} > \tilde{I}_c$, the battery voltage strictly decreases, $d\Delta\tilde{\phi}/d\tilde{c} < 0$, and the homogeneous state is linearly stable for all compositions. The critical current $\tilde{I}_c$ and corresponding spinodal concentration $\tilde{c}_c$ satisfy $\partial\Delta\tilde{\phi}(\tilde{c},\tilde{I})/\partial\tilde{c} = \partial^2\Delta\tilde{\phi}(\tilde{c},\tilde{I})/\partial\tilde{c}^2 = 0$. Using $\Omega = 0.183\,eV$ for LiFePO$_4$ (fitted to the room-temperature miscibility gap), the upper bound for the critical current is $\tilde{I}_c \approx 1.9$, as illustrated in Fig. 3. Far into the solid-solution regime, where $\tilde{I} \gg \tilde{J}_0$, the voltage has a simple Tafel form $\Delta\tilde{\phi} \sim 2\ln\left[(1-\tilde{c})/\tilde{I}\right]$.

Unlike equilibrium phase diagrams, the non-equilibrium stability diagram (Fig. 3) is traversed from left to right during discharge, since $X = \tilde{c} + \tilde{I}t$. Whenever the homogeneously filling state is unstable, we must also determine whether phase separation has enough *time* to occur before the particle becomes full. We suggest the term "quasi-solid solution" to describe a non-equilibrium system that is unstable, but lacks the time to fully phase separate.

Performing a linear stability analysis of Eqs. (5)-(8) around a homogeneous base state $\tilde{c} + \tilde{I}t$, the dimensionless growth rate $\tilde{s} = s\tau_0$ for a perturbation of wavenumber $\tilde{k}$ has the general form:

$$\tilde{s}(\tilde{k},\tilde{c},\tilde{I}) = -\left(\tilde{\mu}'_h(\tilde{c}) + \tilde{K}\tilde{k}^2\right)\sqrt{\left(\frac{\tilde{I}}{2}\right)^2 + \left(\overline{J}_0(\tilde{c})\right)^2} + \left(\frac{\overline{J}'_0(\tilde{c})}{\overline{J}_0(\tilde{c})} + \frac{1}{2}\tilde{K}\tilde{k}^2\right)\tilde{I} \qquad (10)$$

where $\tilde{\mu}_h(\tilde{c})$ and $\overline{J}_0(\tilde{c})$ are for homogeneous states and primes denote $d/d\tilde{c}$ (see supporting information for derivation). The smallest allowable wavenumber is the most unstable, which selects $\tilde{k} = 2\pi$ if boundary effects are neglected. For Eqs. (6)-(7) in one dimension, the marginal stability curve $\tilde{s}_{max}(\tilde{c},\tilde{I}) = \tilde{s}(2\pi,\tilde{c},\tilde{I}) = 0$ is plotted in Fig. 3 (see Eq. (S24)). Requiring that the instability growth time $\tau_I(c,I) = \tau_0/\tilde{s}_{max}(\tilde{c},\tilde{I})$ be greater than the time to fill the particle $\tau_f = Ne/I$ (where $N=N_AN_H$ is the total number of sites) yields a dimensionless relation $\tilde{s}_{max}(\tilde{c},\tilde{I}) < \tilde{I}$, which approximates the conditions for a quasi-solid solution (above the dashed curve in Fig. 3).

The preceding analysis neglects the nonlinear growth of the instability, the statistical occurrence of fluctuations (which tend to excite shorter, less unstable wavelengths), and other effects such as geometry, elasticity, and the amount and configuration of interfaces, all of which can further suppress phase separation significantly. The threshold between quasi-solid solution and phase separation (dotted curve in Fig. 3) thus lies well below the linear instability prediction (dashed curve). To better understand the transition current $\tilde{I}_s$ beyond which phase separation is suppressed, we make a simple scaling argument based on the creation of $N_w$ intercalation waves. If the increased overpotential for the reduced active area $N_w\tilde{\lambda}$ is larger than the voltage gain from homogeneous reaction, $\left|\Delta\tilde{\phi}_{min}\right| \sim \tilde{\Omega}/4$, then a quasi-solid solution is preserved, which implies $\tilde{I}_s \sim N_w\tilde{\lambda}\tilde{\Omega}/4$ (see supporting information). For two waves, we find $\tilde{I}_s \approx 0.2$ with corresponding overpotential $\eta_s \approx 40\text{mV}$, consistent with the simulation results. Typical overpotentials are much larger, so we conclude that phase separation is suppressed in nano-LiFePO$_4$ during normal battery operation.

*Simulations of battery discharge.* – To test our analytical predictions, we simulate reaction-limited lithium insertion in a Li$_x$FePO$_4$ nanoparticle for a range of applied currents and relate phase-separation kinetics to the transient battery voltage (Figs. 4, 5). Movies are available in the supporting information. We solve Eqs. (5)-(8) numerically using an explicit finite-difference method, adjusting $\Delta\tilde{\phi}$ at each time step to maintain the constant current. Langevin noise with a variance scaled to $\tilde{J}_0$ is added to the concentration variable to account for thermal fluctuations.[42] We simulate a particle of length $L$=100nm with a phase boundary thickness [9] of $\lambda$=5nm at room temperature ($V_sK$=0.684 eV·nm$^2$, comparable to Tang *et al.* [20]).

First, we simulate battery discharge for a very small current, $\tilde{I} = 0.01 \ll \tilde{I}_s$, well below the critical current for phase separation (Fig. 3). In the absence of nucleation, the system fills to the equilibrium spinodal point and noise triggers spontaneous phase separation. Two intercalation waves then propagate through the crystal (Fig. 4b). At the spinodal composition, the voltage (Fig. 4a) suddenly jumps to the equilibrium plateau, with a slight overpotential due to a kinetic "wave resistance", inversely proportional to the active interfacial area. When one wave annihilates at the left side facet, there is a small bump in the voltage at constant current due to interfacial energy released when the phase boundary is removed. The overpotential then doubles (since two waves are replaced by one), until the second wave reaches the right side facet, producing another voltage bump and leaving the particle full of lithium. Although the bumps overshoot the theoretical equilibrium plateau at $V^\Theta$, the overpotential relative to the Nernst equilibrium potential $V^\Theta - \mu/e$ remains negative.

Next we consider larger currents with suppressed phase separation. For $\tilde{I} = 0.25 > \tilde{I}_s$, the system behaves as a quasi-solid solution. Linear instability of the concentration profile occurs, but it lacks the time to grow before the particle becomes full (Fig. 4c). The predicted voltage rise during partial instability is unusual for a battery material, but closely resembles some previously unexplained data for reaction-limited nano-LiFePO$_4$ (See 20C curve of Fig. 3a in Ref. 4). For $\tilde{I} = 2 > \tilde{I}_c$, the homogeneous state remains stable throughout the battery discharge (Fig. 4d). It is noteworthy that the voltage is relatively flat for this case (Fig. 4a), which could be misinterpreted as a sign of phase separation. The model also predicts a large overpotential at high filling due to the excluded volume effect, which could be misinterpreted as concentration polarization (diffusion limitation).

Experimental observations suggest that the lithiated phase perfectly wets the inactive side facets of the particle,[10] and so we repeat the simulation with wetting boundary conditions[43] that permit heterogeneous nucleation on the particle surface (Fig. 5). At an small current, $\tilde{I} = 0.01 < \tilde{I}_s$, the wetted boundaries trigger phase separation close to the miscibility gap, and the particle fills at constant voltage by two waves nucleated at the edge. For a larger current, $\tilde{I} = 0.5 > \tilde{I}_s$, although the lithiated phase still wets the boundaries, the waves hardly propagate before the bulk fills like a solid solution (Fig. 5c). Interestingly, the voltage curves show no difference between spinodal decomposition and wetting if $\tilde{I} > 1$.

*Discussion.* – The key fitting parameter in our model is the local exchange current density of LiFePO$_4$ on the active (010) facet, but it is difficult to measure and can vary by orders of magnitude due to the particle size distribution, agglomeration and the surface chemistry. Recent experiments estimate 0.1mA/cm$^2$ and 1.5mA/cm$^2$ over the geometrical area of the electrode,[14,15] not the active surface area of LiFePO$_4$ particles as defined in Eq. (8). If the reacting area is 100 times the geometrical area, then $\tilde{I}$=2 is comparable to 0.3C-rate discharge for the electrode in Ref. 14 and 4.5C-rate for Ref. 15 (Sample A). Regardless of the precise current values, however, an important qualitative feature to note is that the transition from discharge curves with a wide, flat voltage plateau at low current to smooth decay of voltage above the critical current,[14,15] which our model suggests is due to the suppression of phase separation. Similar behavior is seen in ultrafine, coated nano-LiFePO$_4$ powders,[4] where the exchange current density is of order 10mA/cm$^2$, which makes the critical current $\tilde{I} = 2$ comparable to 30C or even higher. Another important observation in this nano-LiFePO$_4$ system is the fact that the 20C discharge curve overshoots and then overlaps the 30C curve after a surprising voltage increase (negative differential capacitance) near the theoretically predicted critical current. Our model suggests that this is a sign of unstable quasi-solid solution behavior. A quantitative fit of the data, however, would require modeling additional effects, such as many-particle interactions, electrolyte diffusion, Ohmic losses, and elastic coherency strain.[34]

It is important to emphasize that there are no measurements of single nanoparticle discharge to test our theory, although a kinetic difference between phase separation in bulk and nano-LiFePO$_4$ has recently been observed.[24] Existing data for composite porous electrodes is complicated by many particle interactions, which have been shown to mask complex phase transformation dynamics.[44] In particular, nearly flat voltage plateaus at very low currents that have long been attributed to single-particle phase separation are likely signs of discrete, one-by-one filling of many particles. For the same reason, the sudden jumps and non-monotonic voltage signatures in our single-particle simulations would be difficult to observe in a macroscopic composite electrode. Moreover, an apparent voltage plateau can mask *in situ* homogeneous intercalation in discharging batteries, even though *ex situ* two-phase coexistence has been detected in electrochemically lithiated particles.[10] As shown in Fig. S3, when a large discharging current is turned off at $X$=0.6 in our simulations, a homogeneous $Li_{0.6}FePO_4$ particle will relax to an equilibrium two-phase system of Li-rich and Li-poor stripes, presumably before the sample can be observed *ex situ*.

*Conclusion.* – In their seminal paper introducing $LiFePO_4$ cathodes, Padhi, Nanjundaswamy, and Goodenough[1] concluded that "the material is very good for *low-power* applications" but "at higher current densities there is a reversible decrease in capacity that … is associated with the movement of a two-phase interface". Ironically, subsequent advances in carbon coating,[45,46] size reduction,[47,48] cation doping,[49] and optimized synthesis methods[50] have made $LiFePO_4$ the most popular cathode material for *high-power* applications. This incredible reversal of fortune has been attributed to changes in material properties below the 100 nm scale, such as enhanced diffusivity[8] and shrinking of the equilibrium miscibility gap,[51] but until now this has been difficult to reconcile with the well documented phase separation behavior near open circuit conditions.[1,9-11]

Our theory suggests that one key to the high rate capability of nano-$LiFePO_4$ is that *phase separation is dynamically suppressed* during normal battery operation. Due to reaction limitation in nanoparticles, the surface overpotential easily exceeds the solid-solution voltage barrier and thus removes the thermodynamic driving force for phase separation, once the current becomes comparable to the exchange current. This contradicts existing models for $LiFePO_4$, which assume artificial phase boundaries and neglect phase separation dynamics. Only at very small currents in large particles should phase separation play a major role, which seems consistent with the improved interpretation of galvanostatic intermittent titration data using an empirical model with a moving phase boundary.[13]

Our results have surprising implications for battery design and utilization. The theory predicts that increasing the interfacial resistance at the active surface tends to further suppress phase transformation, thereby increasing the available active area for intercalation. This leads to the counterintuitive conclusion that slowing the surface reaction could be beneficial for battery performance, which may explain the improved discharge rate capability of $LiFePO_4$ nanoparticles with thin (5nm) phosphate glass coatings.[4] Since phase separation causes mechanical deformation due to lattice-mismatch strain, its suppression would also reduce stresses in the crystal that can contribute to capacity fade, e.g. by point defect formation[8] or iron leaching.[52] In other words, very slow discharge could actually *reduce* the cycle life of the battery.

Our arguments are quite general and may only be reinforced by including more physics in the model, as suggested by ongoing work in our group. Elastic coherency strain further suppresses instability of the homogeneously filling state. Including size-dependent diffusion would help to understand the transition to phase separating behavior in larger particles, but does not affect nanoparticle behavior. Finally, including electrolyte depletion in a composite electrode would allow us to make connections with experimental data, by extending porous electrode theory based on quasi-solid solution nanoparticles.

Beyond Li-ion batteries, our theory involves new concepts in non-equilibrium thermodynamics, which may have other applications. Phase transformations are typically studied in closed or periodic bulk systems, close to equilibrium.[18] Our theory describes the suppression of phase transformations in a driven, open system, away from equilibrium. This leads to the notion of an unstable quasi-solid solution, which exists only for certain currents in a non-equilibrium extension of the phase diagram (Fig. 3). Other examples of this general phenomenon may include transient clustering in electrophoretic displays[53] or delayed shock-induced phase transformations in solids.[54]

**Acknowledgement.** This work was supported by the National Science Foundation under Contracts DMS-0842504 and DMS-0948071 and by a seed grant from the MIT Energy Initiative. P. B. also acknowledges the support of a Chinese Government Scholarship, File No: 2009621147.

**Supporting Information Available.** Regular solution model, modified Butler-Volmer equation, instability growth rate, 1D wave velocity, relaxation behavior and five simulation movies.

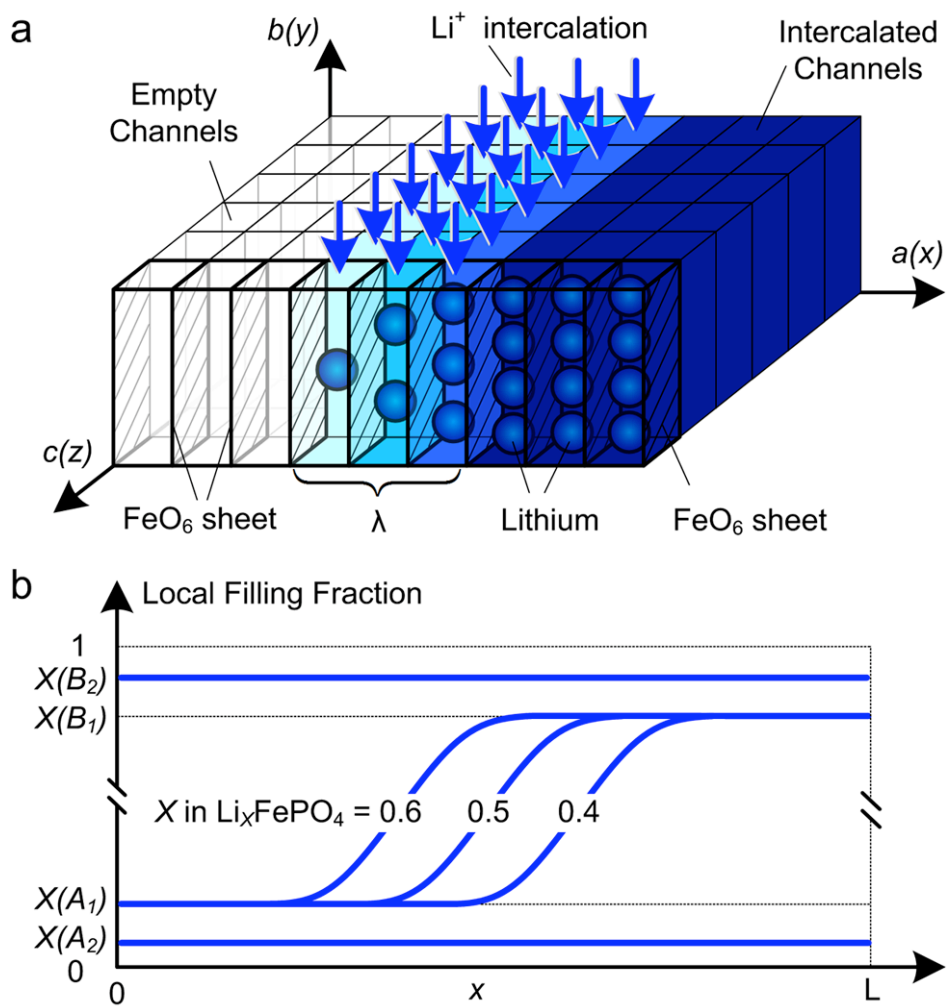

**Figure 1.** Schematic model of a Li$_X$FePO$_4$ nanoparticle at low overpotential. (FeO$_6$ and PO$_4$ structures are not shown in order to highlight inserted Li.) (a) Lithium ions are inserted into the particle (blue arrows) from the active (010) facet with fast diffusion and no phase separation in the depth ($y$) direction, forming a phase boundary of thickness $\lambda$ between full and empty channels. (b) The resulting 1D concentration profile (local filling fraction) transverse to the FePO$_4$ planes for a particle of size L. The average compositions $X$=0.4, 0.5 and 0.6 reflect mixtures of coexisting Li-poor and Li-rich phases, $X(A_1)$ and $X(B_1)$ respectively, which vary as the phase boundary moves during charging or discharging. Homogeneous profiles at high and low concentrations, $X(A_2)$ and $X(B_2)$ respectively (see Fig. 2), are also shown.

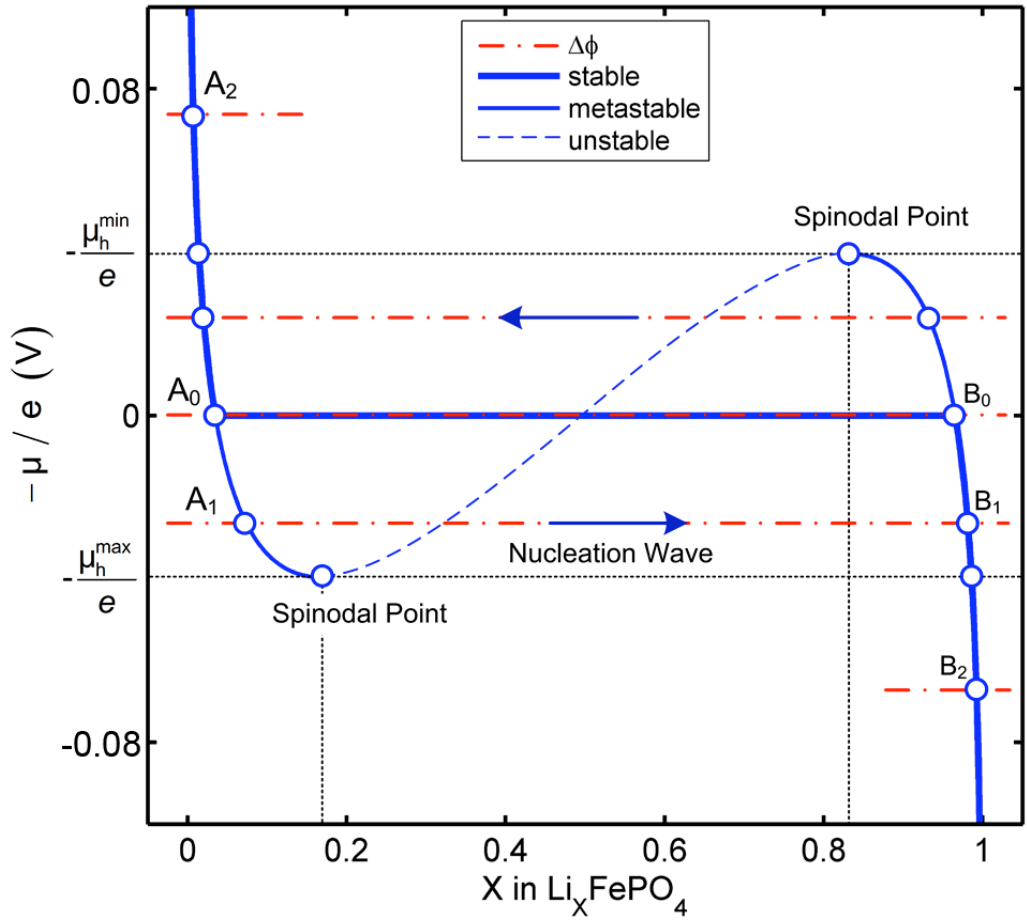

**Figure 2.** Plot of the diffusional chemical potential against the filling fraction of the particle at equilibrium. The thick solid curves are the equilibrium chemical potential in a phase-separating system. If phase-separation is suppressed, the chemical potential deviates from the voltage plateau instead following the metastable and unstable pathway. Dash-dot (red) lines are different $\Delta\phi$, and small circles indicate equilibrium states at different applied voltages.

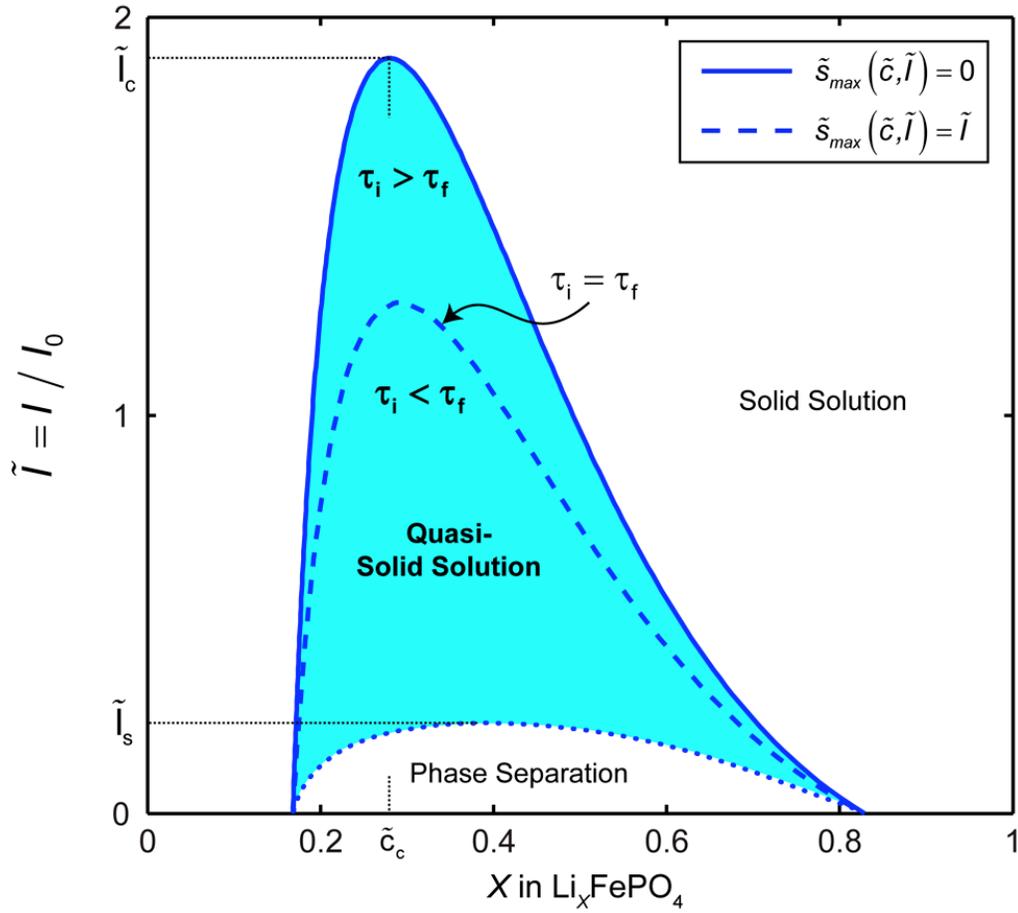

**Figure 3.** Dependence of spinodal region on the applied current. The solid curve gives the marginal stability of the system under applied current, and the dashed curve is the boundary between phase separation and quasi-solid solution according to linear stability analysis. In practice, the boundary where phase separation is observed is expected to be significantly below the linear stability limits, e.g. the dotted curve is inferred from numerical simulations and scaling arguments.

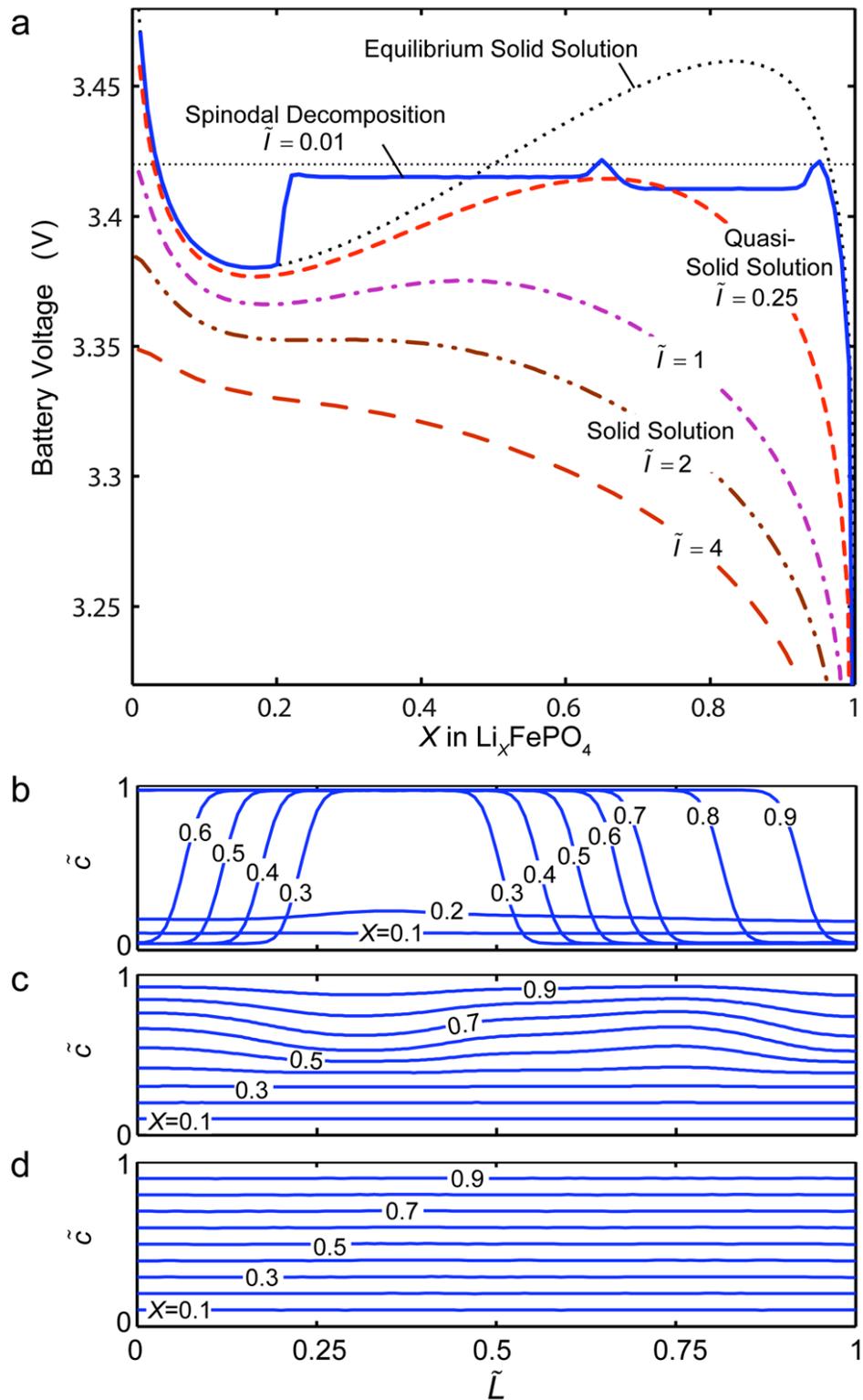

**Figure 4.** Numerical simulation of phase transformation via spinodal decomposition. (a) Voltage responses at different constant currents. Concentration evolution of (b) spinodal decomposition at $\tilde{I}=0.01$, (c) quasi-solid solution at $\tilde{I}=0.25$, and (d) solid solution at $\tilde{I}=2$. Labels on the concentration curves indicate the mean filling fraction of the particle.

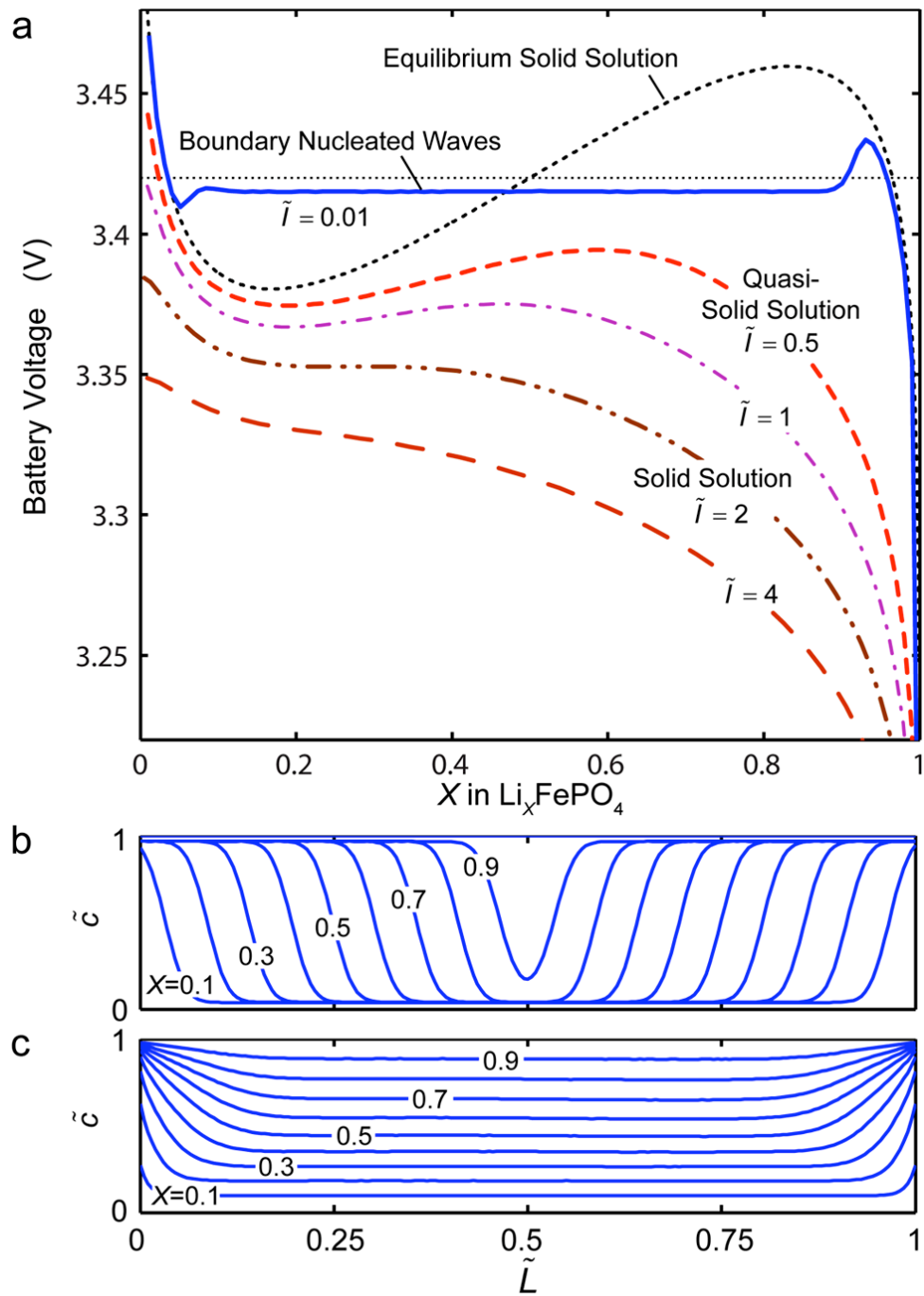

**Figure 5.** Numerical simulation of phase transformation triggered by wetting of the particle boundary. (a) Voltage responses at different constant currents. Concentration evolution of (b) waves at $\tilde{I}$=0.01, and (c) quasi-solid solution at $\tilde{I}$=0.5.

# Supporting Information

1. **Regular Solution Model**

$$g_{\text{hom}} = \Omega \tilde{c}(1-\tilde{c}) + 2k_B T \tilde{c} \ln \tilde{c} + 2k_B T (1-\tilde{c}) \ln(1-\tilde{c}) \tag{S1}$$

$$\mu_{\text{hom}} = \Omega(1-2\tilde{c}) + 2k_B T \ln\left(\frac{\tilde{c}}{1-\tilde{c}}\right) \tag{S2}$$

Fig. S1a shows a free energy curve with $\Omega = 0.183$ eV, which gives a broad miscibility gap of 0.035~0.965 at room-temperature, in agreement with the LiFePO$_4$ phase diagram (*1*). Fig. S1b shows the corresponding homogeneous diffusional chemical potential with the horizontal line indicating the equilibrium voltage plateau. Note that the vertical axis is negative since battery voltage is $V_B = V^{\ominus} + \eta - \mu/e$. Fig. S1 also demonstrates how increasing temperature affects the free energy and chemical potential via the entropic term, which is consistent with experimental findings of solid solution behavior at high temperature (*2*), where *in situ* XRD has demonstrated the signature peak of LiFePO$_4$ continuously moving to the position of FePO$_4$. One can easily observe in Fig. S1 that increasing temperature causes the energy barrier to decrease, and it eventually disappears at 350 °C. There is no common tangent, and thus only one phase exists at equilibrium.

2. **Derivation of the Modified Butler Volmer Equation**

We can modify the Butler-Volmer equation for LiFePO$_4$ electrodes with concentrated electrolyte by simply replacing concentrations with activities to account for the short-range ionic interactions. In *dynamic* situations though, the effect of the concentrated solution on the transition state must also be considered. Excess chemical potential, which drives the lithium away from the base energy level (ideal solution), is used to model the rate equation. In a general first-order electrode reaction, $O^+ + e^- \rightleftharpoons R$, we can write the overall electrochemical potential for the reduced and the oxidized states as:

$$\mu_O = \mu_O^{id} + \mu_O^{ex} = k_B T \ln c_O + \mu_O^{ex} = k_B T \ln a_O - e\Delta\phi \tag{S3}$$

$$\mu_R = \mu_R^{id} + \mu_R^{ex} = k_B T \ln c_R + \mu_R^{ex} = k_B T \ln a_R \tag{S4}$$

Where $\mu_i$, $c_i$, $a_i$, *(i=O,R)*, are the electrochemical potential, ionic concentration, and activities for reduced (or oxidized) state. Both electrochemical potentials consist of the ideal part $\mu_i^{id}$, and excess part $\mu_i^{ex}$. The last term in Eq. (S3) accounts for the electrostatic potentials of ions in solution and electrons in the metallic phase with $\Delta\phi=(\phi_e-\phi)$. $k_B$, $T$, $e$ are Boltzmann's constant, temperature and elementary charge respectively.

We further postulate the excess chemical potential for the activated state as:

$$\mu_A^{ex} = k_B T \ln \gamma_A - (1-\alpha) e \Delta\phi \tag{S5}$$

Where $\gamma_A$ is the activity coefficient of the activated lithium ions and $\alpha$ is the charge transfer coefficient of the redox reaction, usually taken as 0.5 for symmetric reactions.

By using transition state theory for concentrated solutions (3), we relate the chemical potentials to the local current density $J$. The forward and backward reaction current density are modeled as:

$$J_{\rightarrow} = \frac{e}{A_s \tau_0} c_O \exp\left[-(\mu_A^{ex} - \mu_O^{ex})/k_B T\right] \tag{S6}$$

$$J_{\leftarrow} = \frac{e}{A_s \tau_0} c_R \exp\left[-(\mu_A^{ex} - \mu_R^{ex})/k_B T\right] \tag{S7}$$

Thus, the overall reaction rate is:

$$\begin{aligned}
J &= J_{\rightarrow} - J_{\leftarrow} \\
&= \frac{e}{A_s \tau_0} \left\{ \exp\left[-(\mu_A^{ex} - \mu_O)/k_B T\right] - \exp\left[-(\mu_A^{ex} - \mu_R)/k_B T\right] \right\} \\
&= \frac{e}{A_s \tau_0 \gamma_A} \left\{ a_O \exp\left[-\alpha e \Delta\phi / k_B T\right] - a_R \exp\left[(1-\alpha) e \Delta\phi / k_B T\right] \right\}
\end{aligned} \tag{S8}$$

Recall that the interfacial voltage is $\Delta\phi = \Delta\phi_{eq} + \eta$, and $\Delta\phi_{eq}$ is the electrostatic potential difference at equilibrium given by Nernst equation:

$$\Delta\phi_{eq} = \frac{k_B T}{e} \ln \frac{a_O}{a_R} \tag{S9}$$

Substituting Eq. (S9) into Eq. (S8) yields:

$$J = \frac{e\, a_O^{1-\alpha} a_R^{\alpha}}{A_s \tau_0 \gamma_A} \left( \exp\left[-\alpha \frac{e\eta}{k_B T}\right] - \exp\left[(1-\alpha) \frac{e\eta}{k_B T}\right] \right) \tag{S10}$$

This is the local current density for lithium insertion process. It reduces to the standard Butler-Volmer equation for $\gamma_- = \gamma_+ = \gamma_A = 1$.

3. **Instability Growth Rate**

Here we present a derivation the linear stability of the reduced model (Eq. 10 of the main text). For simplicity, tilde notation for dimensionless variables has been dropped even though variables in the analysis are assumed to be dimensionless. Assuming $\alpha = 0.5$, the evolution equations for the reduced model are:

$$\frac{\partial c}{\partial t} = -2J_0(c, \nabla^2 c) \sinh\left(\frac{\eta(c, \nabla^2 c)}{2}\right) \tag{S11}$$

$$J_0(c, \nabla^2 c) = (1-c)e^{\mu(c,\nabla^2 c)/2} \tag{S12}$$

$$\eta(c, \nabla^2 c) = \Delta\phi + \mu(c, \nabla^2 c) \tag{S13}$$

$$I = \int_0^1 \frac{\partial c}{\partial t} dx \tag{S14}$$

To perform the linear stability analysis, we examine the growth of small perturbations about a homogeneous, constant current base state. We let:

$$c = c_0 + It + v \tag{S15}$$

where $c_0$ is a constant composition, $I$ is a constant current, $c_h = c_0 + It$ is the homogeneous base state, and $v = e^{ikx}e^{st}$ is a small perturbation. The growth factor $s$ determines if a perturbation with wave number $k$ will be amplified. The derivatives of $c$ are:

$$\frac{\partial c}{\partial t} = I + sv \qquad \frac{\partial c}{\partial x} = ikv \tag{S16}$$

For a homogeneous system, $c = c_h$ and $\nabla^2 c = 0$, and Eq. (S14) reduces to:

$$I = -2\bar{J}_0 \sinh\left(\frac{\bar{\eta}}{2}\right) \tag{S17}$$

Bar notation here and throughout the derivation indicates evaluation at the homogeneous state $c_h$.

Next we linearize the evolution equations and solve for the growth factor. The Taylor expansion of a function $f(c, \nabla^2 c)$ about a homogeneous state $\bar{f} = f(c_h, 0)$ is:

$$f(c, \nabla^2 c) = \bar{f} + \overline{\left(\frac{\partial f}{\partial c}\right)}(c - c_h) + \overline{\left(\frac{\partial f}{\partial \nabla^2 c}\right)}\nabla^2 c + \ldots \tag{S18}$$

Keeping only the linear terms and substituting Eq. (S15) yields a linear approximation to $f$:

$$f(c, \nabla^2 c) \approx \bar{f} + \overline{\left(\frac{\partial f}{\partial c}\right)}v + \overline{\left(\frac{\partial f}{\partial \nabla^2 c}\right)}(-k^2 v) \tag{S19}$$

Next we use Eq. (S15) and Eq. (S19) to linearize the evolution equation (Eq. (S11)):

$$\frac{\partial c}{\partial t} = I + sv = -2\bar{J}_0 \sinh\left(\frac{\bar{\eta}}{2}\right)$$

$$-\left(\bar{J}_0\bar{\mu}'\cosh\left(\frac{\bar{\eta}}{2}\right) + 2\bar{J}_0'\sinh\left(\frac{\bar{\eta}}{2}\right)\right)v \quad (S20)$$

$$+\left(\bar{J}_0 K \cosh\left(\frac{\bar{\eta}}{2}\right) + \bar{J}_0 K \sinh\left(\frac{\bar{\eta}}{2}\right)\right)(-k^2 v)$$

Solving for $s$ and simplifying with the use of Eq. (S17) produces:

$$s = -\left(\bar{\mu}' + Kk^2\right)\bar{J}_0 \cosh\left(\frac{\bar{\eta}}{2}\right) + \left(\frac{\bar{J}_0'}{\bar{J}_0} + \frac{1}{2}Kk^2\right)I \quad (S21)$$

The cosh term in Eq. (S21) can be simplified using the hyperbolic trig identity $\cosh^2(z) = 1 + \sinh^2(z)$ and Eq. (S17):

$$\bar{J}_0 \cosh\left(\frac{\bar{\eta}}{2}\right) = \sqrt{\bar{J}_0^2 + \bar{J}_0^2 \sinh^2\left(\frac{\bar{\eta}}{2}\right)} = \sqrt{\bar{J}_0^2 + \frac{I^2}{4}} \quad (S22)$$

Substitution of the above expression into Eq. (S21) yields the growth rate:

$$s(k, c, I) = -\left(\bar{\mu}' + Kk^2\right)\sqrt{\bar{J}_0^2 + \frac{I^2}{4}} + \left(\frac{\bar{J}_0'}{\bar{J}_0} + \frac{1}{2}Kk^2\right)I \quad (S23)$$

For our model (6)-(7) in one dimension, the general formula (10) for the growth rate of the most unstable mode ($\tilde{k} = 2\pi$) can be evaluated in terms of $\tilde{\delta} = 1 - 2\tilde{c}$:

$$\tilde{s}_{\max}(\tilde{c}, \tilde{I}) = \left(\tilde{\Omega} - \frac{4}{1-\tilde{\delta}^2} - 2\tilde{K}\pi^2\right)\sqrt{\tilde{I}^2 + \left(1-\tilde{\delta}\right)^2 e^{\tilde{\Omega}\tilde{\delta}}} + \left(\frac{2}{1-\tilde{\delta}} - \tilde{\Omega} + 2\tilde{K}\pi^2\right)\tilde{I} \quad (S24)$$

4. **1D Wave Velocity**

Following SCB (*4*), it can be shown that Eq. (5) has traveling-wave solutions of the form $\tilde{c}(\tilde{x}, \tilde{t}) = f(\tilde{x} - \tilde{v}\tilde{t})$, corresponding to a moving phase boundary. We can get the current carried by a single wave as:

$$\tilde{I}_w = \int_0^1 \frac{\partial \tilde{c}}{\partial \tilde{t}} d\tilde{x} = -\tilde{v}\int_{-\infty}^{+\infty} \frac{\partial f}{\partial \zeta} d\zeta = -\tilde{v}\Delta f = -\tilde{v}\Delta\tilde{c} \quad (S25)$$

where, $\Delta\tilde{c} \approx 1$ is the width of the shifted miscibility gap, e.g. length of $A_1B_1$ in Fig. 2.

In the case of phase separation, whether the wave starts by instability or nucleation, the late stage of phase separation leads to $N_w$ intercalation waves with smaller active area by a factor of $N_w\tilde{\lambda}$. The overpotential increase for a single wave (i.e. within the phase boundary) is $\tilde{\eta}_w = \tilde{\mu}_h(\tilde{c}) - \tilde{K}\partial^2\tilde{c}/\partial\tilde{x}^2 + \Delta\tilde{\phi}$. Keep in mind that in the inactive region far away from the diffuse phase boundary, we have $\partial^2\tilde{c}/\partial\tilde{x}^2 \to 0$, $\tilde{\mu}_h(\tilde{c}) \to \Delta\tilde{\phi}$ and therefore $\tilde{\eta} \to 0$ (negligible reactions).

In the active region of the diffuse phase boundary, the overpotential is nonzero and can be estimated as follows. For small currents, the phase boundary maintains a quasi-equilibrium concentration profile, given by $\tilde{\mu}_h(\tilde{c}) \approx \tilde{K}\partial^2\tilde{c}/\partial\tilde{x}^2$, which implies $\tilde{\eta}_w \approx \Delta\tilde{\phi}$ in the active region. The wave velocity can then be estimated as:

$$\begin{aligned}
\tilde{v} \approx \tilde{I}_w &= \int_0^1 \frac{\partial \tilde{c}}{\partial \tilde{t}} d\tilde{x} \\
&\approx -2\int_{\tilde{x}_0(t)-\tilde{\lambda}/2}^{\tilde{x}_0(t)+\tilde{\lambda}/2} \tilde{J}_0 \sinh\left(\frac{\tilde{\eta}_w}{2}\right) d\tilde{x} \\
&= -2\int_{\tilde{x}_0(t)-\tilde{\lambda}/2}^{\tilde{x}_0(t)+\tilde{\lambda}/2} (1-\tilde{c})\exp\left(\tilde{\eta}_w - \Delta\tilde{\phi}\right)\sinh\left(\frac{\tilde{\eta}_w}{2}\right) d\tilde{x} \qquad (S26) \\
&\approx -2\int_{\tilde{x}_0(t)-\tilde{\lambda}/2}^{\tilde{x}_0(t)+\tilde{\lambda}/2} (1-\tilde{c}) \cdot 1 \cdot \sinh\left(\frac{\Delta\tilde{\phi}}{2}\right) d\tilde{x} \\
&\approx -\Delta\tilde{\phi} \int_{\tilde{x}_0(t)-\tilde{\lambda}/2}^{\tilde{x}_0(t)+\tilde{\lambda}/2} (1-\tilde{c}) d\tilde{x}
\end{aligned}$$

where $\tilde{x}_0(\tilde{t}) = \tilde{x}_0(0) - \tilde{v}\tilde{t}$ is the (arbitrary) center of the wavefront. Finally, assuming the averaged concentration within the phase boundary is roughly 1/2, Eq. (S26) reduces to:

$$\tilde{v} \approx -\tilde{\lambda}\Delta\tilde{\phi}/2 \qquad (S27)$$

As shown in Fig. S2, in spite of various crude approximations, Eq. (S27) provides a good estimate of the wave velocity at small overpotentials.

For larger currents, the intercalation wave loses its quasi-equilibrium profile and Eq. (S27) no longer holds. Accurate estimation of the velocity requires sophisticated analysis on the interfacial energy of the moving phase boundary. Instead, we can make simple scaling analysis to capture the basic physics. Since the active area, $N_w\tilde{\lambda}$, can be understood physically as the "kinetic conductance", the current carried by the waves can be estimated by Ohm's law as $\tilde{I} \sim -N_w\tilde{\lambda}\Delta\tilde{\phi}$. Estimation of $\tilde{I}_s$ via this simple formula is close to our simulation inference of the threshold, beyond which traveling wave solutions no longer exist and phase separation is suppressed. As shown in Fig. S2, the velocity of the intercalation wave significantly deviates from the prediction of Eq. (S27) (the assumption of quasi-equilibrium) and eventually disappears when $\Delta\tilde{\phi} > \Delta\tilde{\phi}_{ss}/2$.

## 5. Relaxation Behavior

If a system is discharged with a large current and the current is turned off when X=0.6, the homogeneous $Li_{0.6}FePO_4$ particle will relax toward an equilibrium two-phase system. In experiment, this relaxation might occur after the discharge is stopped but before the sample can be observed *ex situ*.

Fig. S3 shows a simulation of a 2D particle relaxation. Starting from the homogeneous state, the system begins to evolve and eventually reaches the phase-separated equilibrium state. Anisotropy in interfacial energy causes the interface to be oriented vertically. Only systems with planar phase boundaries can reach true equilibrium; systems with curved phase boundaries will never equilibrate to $V^\Theta = 3.42V$ but leave a "hysteresis gap". Elastic effects are responsible for the larger hysteresis that is often observed experimentally (*5*).

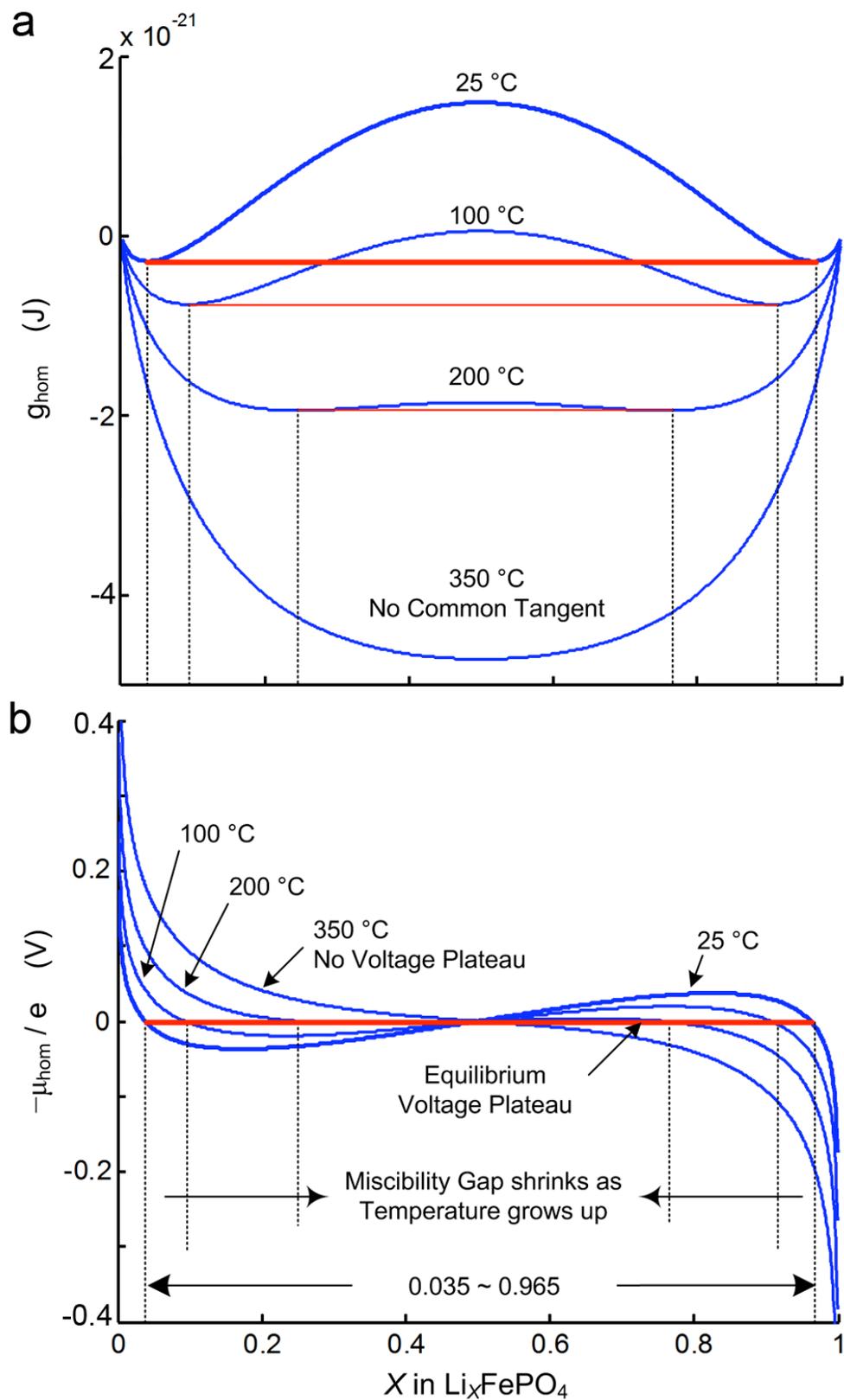

**Figure S1.** Free energy and chemical potential for the regular solution model.

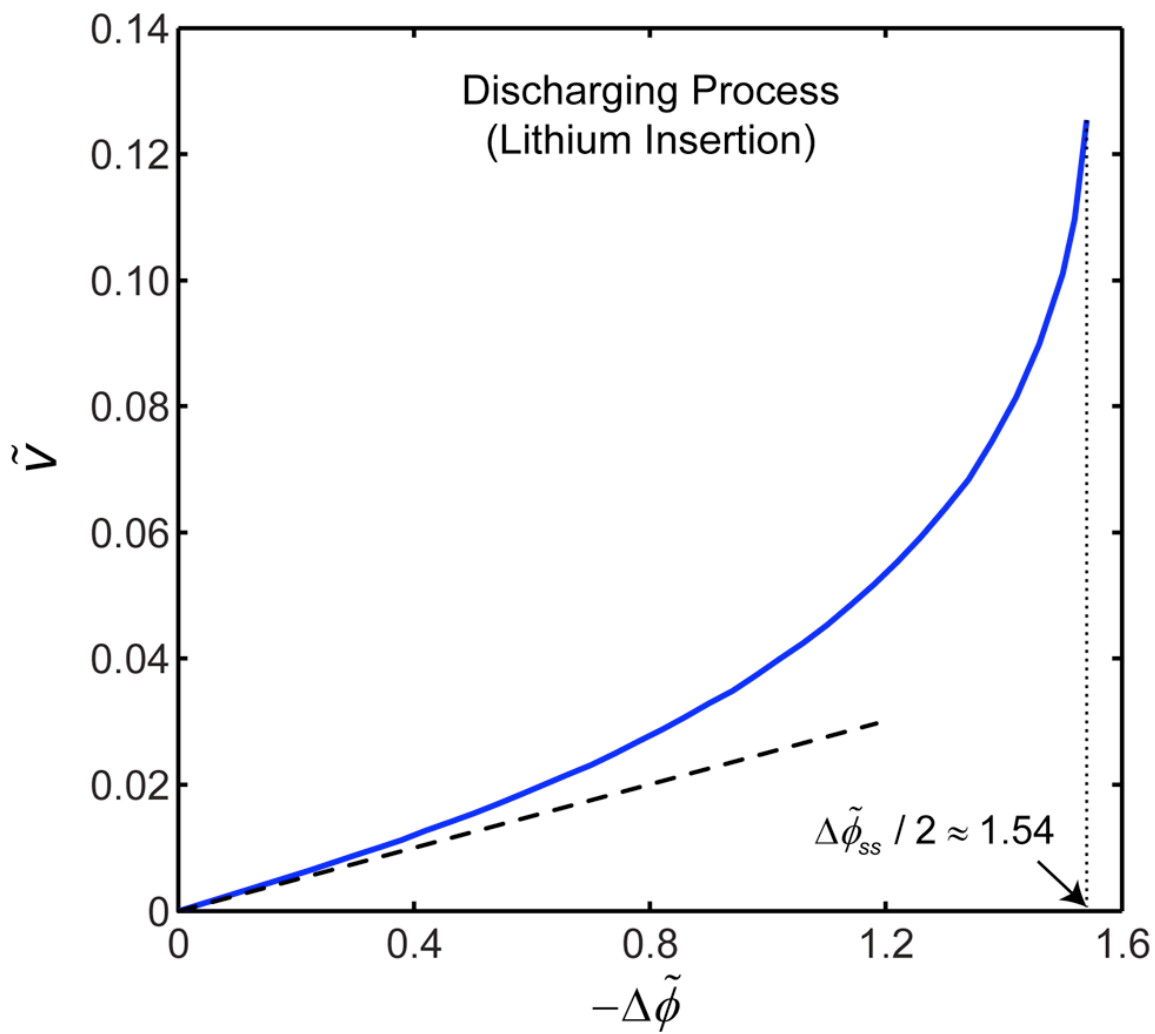

**Figure S2.** Dependence of intercalation wave velocity on applied voltage. Solid curve is estimated from simulation. Wetting boundaries are used to trigger a single traveling wave. Dashed line comes from Eq. (S27).

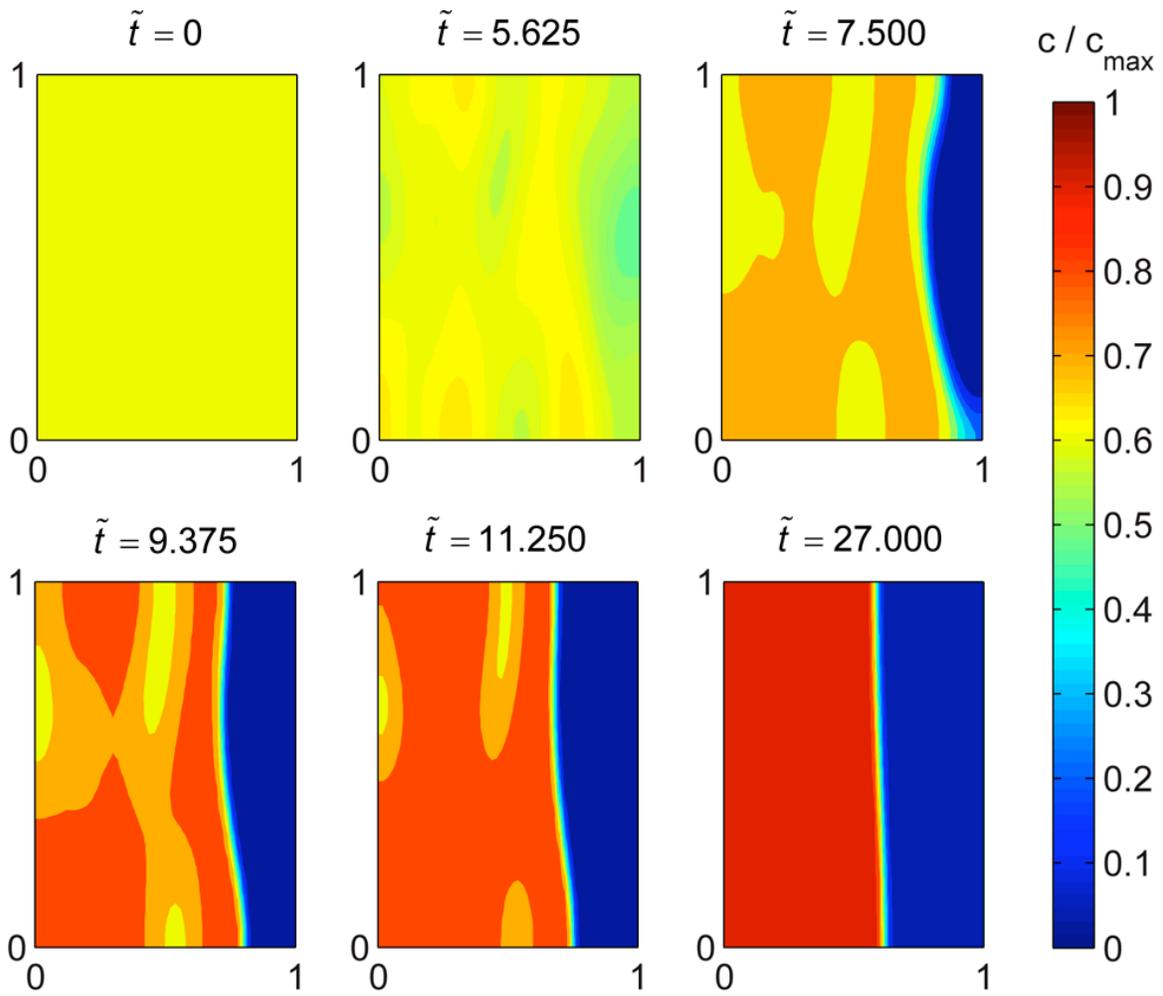

**Figure S3.** Numerical simulation of phase separation in a Li$_{0.6}$FePO$_4$ particle (top surface of the particle in Fig.1a) during relaxation from a solid solution. $\tilde{t} = t/\tau_0$ is the dimensionless time, scaled to the time constant of surface reaction. The Cahn-Hilliard $K$ is extended to two dimensions with $K_z \gg K_x$ in order to mimic the elastic effects that align the interface. Since phase separation is triggered by noise, phase boundaries with different orientations and geometries are expected in each simulation.

27